\documentclass[12pt]{article}
\usepackage{graphicx}
\usepackage{amsmath}
\usepackage{color}
\input psfig.sty

\newcommand{\lsim}{\mathrel{\vcenter{\hbox{$<$}\nointerlineskip\hbox{$\sim$}}}}

\begin{document}

\begin{titlepage}

\title{\bf Evolution of the Universe with \\ Flat Extra Dimensions\\
\vspace{0.5cm}}

\author{Je-An Gu\thanks{%
E-mail address: jagu@phys.ntu.edu.tw}\ , \ W-Y.~P.~Hwang\thanks{%
E-mail address: wyhwang@phys.ntu.edu.tw}\ \ \ and \ Jr-Wei Tsai \\
{\small Department of Physics, National Taiwan University, Taipei
106, Taiwan, R.O.C.}
\medskip
}

\date{\small \today}

\maketitle

\begin{abstract}
Evolution of a universe with homogeneous extra dimensions is
studied with the benefit of a well-chosen parameter space that
provides a systematic, useful, and convenient way for analysis. In
this model we find a natural evolution pattern that entails not
only stable extra dimensions in the radiation-dominated era,
thereby preserving essential predictions in the standard
cosmology, but also the present accelerating expansion while
satisfying the limit on the variation of Newtonian gravitational
constant. In this natural evolution pattern the extra dimensions
tend to be stabilized automatically without resorting to
artificial mechanisms in both the radiation-dominated and the
matter-dominated era, as a wonderful feature for building models
with extra dimensions. In addition, the naturalness of this
evolution pattern that guarantees the late-time accelerating
expansion of a matter-dominated universe presents a solution to
the coincidence problem: why the accelerating phase starts at the
present epoch. The feasibility of this evolution pattern for
describing our universe is discussed.
\end{abstract}


\thispagestyle{empty}

\end{titlepage}

\section{Introduction}

The observations by Supernova Cosmology Project and Supernova
Search Team have suggested in 1998 that the expansion of the
present universe is accelerating
\cite{Perlmutter:1999np,Riess:1998cb}. This conclusion is
reinforced recently in 2003 by WMAP measurements \cite{WMAP2003}.
One general conclusion from these measurements and the CMB
observations in recent years
\cite{Sievers:2002tq,Kuo:2002ua,WMAP2003} is that the universe has
the critical density, consisting of $1/3$ of ordinary matter and
$2/3$ of dark energy with a negative pressure \cite{CMB&SN} (such
that $p_{\textsc{x}}/\rho_{\textsc{x}} < -0.78$ \cite{WMAP2003}).
Although this acceleration may be driven by the existence of a
positive cosmological constant (vacuum energy) \cite{Lambda
models}, there remain other interpretations of the accelerating
expansion, such as ``quintessence'' (a slowly evolving scalar
field) \cite{Caldwell:1998ii,ComplexQ} and the presence of extra
dimensions \cite{Gu:2001ni,Gu:2002mz}.

The presence of extra dimensions is required in various theories
beyond the standard model of particle physics, especially in the
theories for unifying gravity and other forces, such as
superstring theory. Extra dimensions should be ``hidden'' (or
``dark'') for consistency with observations. Various scenarios for
``hidden'' extra dimensions have been proposed, for example, a
brane world with large compact extra dimensions in factorizable
geometry proposed by Arkani-Hamed \emph{et al.}
\cite{Arkani-Hamed} (see also \cite{Antoniadis:1990ew}), and a
brane world with noncompact extra dimensions in nonfactorizable
geometry proposed by Randall and Sundrum \cite{Randall&Sundrum}.
In this paper we will employ the simplest scenario: small compact
extra dimensions in factorizable geometry, as introduced in the
Kaluza-Klein (KK) theories \cite{Kaluza&Klein}.

The possibility of generating the accelerating expansion of the
present matter-dominated universe via the evolution of homogeneous
and isotropic extra dimensions is first pointed out by Gu and
Hwang \cite{Gu:2001ni}. Furthermore, the general idea of unifying
dark energy sources (i.e.\ dark matter and dark energy) and dark
geometry (e.g.\ extra dimensions) into one has been sketched by Gu
\cite{Gu:2002mz}, who pointed out that dark geometry (instead of
dark energy) can be an intriguing candidate for generating
accelerating expansion. As indicated by Einstein's general
relativity, mass (energy) and geometry are two faces of one single
nature, therefore it is biased to consider only mass (energy) has
dark part, while believing blindly that all geometry is totally
``visible'' to our poor eyes \cite{Gu:2002mz}.

In this simple scenario making use of a highly symmetric extra
space, there are much fewer free parameters (only two additional
degrees of freedom, the expansion rate and the curvature of the
extra space, in addition to those in the standard cosmological
models without dark energy) so that it is much easier to be ruled
out, compared with the quintessence models, %
by constraints from observations. In particular, an essential
difficulty of this model \cite{Cline:2002mw} stems from the
constraint on the variation of the Newtonian gravitational
constant, which will be produced along with the evolution of extra
dimensions that is the key element for generating accelerating
expansion.

In this paper we will study a more general case in which the
evolution of our universe in various eras, especially the present
accelerating expansion era, is governed not only by the matter
contents (excluding dark energy) therein, but also by the
curvature of the ordinary 3-space and the evolution of extra
dimensions. We will explore the feasibility of this model for
generating accelerating expansion while satisfying observational
constraints, especially the variation of the Newtonian
gravitational constant. %
Through these studies we will also present several nice features
of this model, such as the automatic stabilization of extra
dimensions and the solution to the cosmic coincidence problem
--- why the energy densities of dark energy and dark matter are
comparable now (i.e.\ ``why now'' problem), or, more precisely (if
dark energy is not a necessary ingredient for the accelerating
expansion), why the accelerating phase of our universe starts at
the present
epoch. %

\section{The Higher-dimensional World: Basics}

We consider a $(3+n+1)$-dimensional space-time where $n$ is the
number of extra spatial dimensions. Assuming that both the
three-dimensional ordinary space and the $n$-dimensional extra
space are homogeneous and isotropic, we use two spatial parts of
the Robertson-Walker metric to describe this space-time as
follows:
\begin{equation}
ds^2 = dt^2 - a^2(t) \left( \frac{dr_a^2}{1-k_a r_a^2} + r_a^2 d
\Omega_a^2 \right) - b^2(t) \left( \frac{dr_b^2}{1-k_b r_b^2} +
r_b^2 d \Omega_b^2 \right) , %
\label{unperturbed metric}
\end{equation}
where $a(t)$ and $b(t)$ are scale factors, and $k_a$ and $k_b$
relate to curvatures of the ordinary 3-space and the extra space,
respectively. The value of $r_b$ is set to be within the interval
$[0,1)$ corresponding to the finite size of extra dimensions.
Assuming that the matter content in this higher-dimensional space
is a perfect fluid with the energy-momentum tensor
\begin{equation}
T^{\alpha}_{\; \; \beta}=diag(\bar{\rho},-\bar{p}_a, \ldots ,
-\bar{p}_b , \ldots ) ,
\end{equation}
we can write the Einstein equations \cite{Gu:2001ni,Gu:2002mz} as
\begin{equation}
3 \left[ \left( \frac{\dot{a}}{a}\right)^2 + \frac{k_a}{a^2}
\right] + \frac{n(n-1)}{2} \left[ \left(
\frac{\dot{b}}{b}\right)^2 + \frac{k_b}{b^2} \right] +
3n\frac{\dot{a}}{a} \frac{\dot{b}}{b} = 8 \pi \bar{G} \bar{\rho}
\, , \label{G00 eq with homog ED}
\end{equation}
\begin{equation}
2 \frac{\ddot{a}}{a} + n\frac{\ddot{b}}{b} + \left[ \left(
\frac{\dot{a}}{a} \right)^2 + \frac{k_a}{a^2} \right] +
\frac{n(n-1)}{2} \left[ \left( \frac{\dot{b}}{b}\right)^2 +
\frac{k_b}{b^2} \right] + 2n \frac{\dot{a}}{a} \frac{\dot{b}}{b}
= - 8 \pi \bar{G} \bar{p}_a \, , %
\label{Gii eq with homog ED}
\end{equation}
\begin{eqnarray}
3 \frac{\ddot{a}}{a} + (n-1)\frac{\ddot{b}}{b} + 3 \left[ \left(
\frac{\dot{a}}{a} \right)^2 + \frac{k_a}{a^2} \right] +
\frac{(n-1)(n-2)}{2} \left[ \left( \frac{\dot{b}}{b}\right)^2 +
\frac{k_b}{b^2} \right] && \nonumber \\
+ 3(n-1) \frac{\dot{a}}{a} \frac{\dot{b}}{b}
= - 8 \pi \bar{G} \bar{p}_b \, , && %
\label{Gjj eq with homog ED}
\end{eqnarray}
where $\bar{\rho}$ is and the energy density in the
higher-dimensional world, and $\bar{p}_a$ and $\bar{p}_b$ are the
pressures in the ordinary 3-space and the extra space,
respectively.

The terms involving the scale factor $b$ in Eqs.\ (\ref{G00 eq
with homog ED}) and (\ref{Gii eq with homog ED}) are additional
terms and Eq.\ (\ref{Gjj eq with homog ED}) is an additional
equation coming from extra dimensions. As pointed out by Gu in
\cite{Gu:2002mz}, if extra dimensions exist but we are so biased
that we admit only mass/energy can have dark part while blindly
believing there is no dark geometry (such as extra dimensions) at
all, these additional terms from extra dimensions will
automatically be moved from the left-hand side of the Einstein
equations (describing geometry) to the right-hand side (describing
energy contents) and be treated as some sort of effective dark
energy sources, as indicated in the following equations [from
rearranging Eqs.\ (\ref{G00 eq with homog ED})--(\ref{Gjj eq with
homog ED})]:
\begin{eqnarray}
3 \left[ \left( \frac{\dot{a}}{a}\right)^2 + \frac{k_a}{a^2}
\right] &=& 8 \pi \bar{G} \left( \bar{\rho} + \bar{\rho}_{eff}
\right) \, , \label{G00 eq with eff DE from ED} \\
2 \frac{\ddot{a}}{a} + \left[ \left( \frac{\dot{a}}{a} \right)^2 +
\frac{k_a}{a^2} \right] &=& - 8 \pi \bar{G} \left( \bar{p}_a +
\bar{p}_{a,eff} \right) \, , \label{Gii eq with eff DE from ED} \\
3 \frac{\ddot{a}}{a} + 3 \left[ \left( \frac{\dot{a}}{a} \right)^2
+ \frac{k_a}{a^2} \right] &=& - 8 \pi \bar{G} \left( \bar{p}_b +
\bar{p}_{b,eff} \right) \, , \label{Gjj eq with eff DE from ED}
\end{eqnarray}
where
\begin{eqnarray}
\bar{\rho}_{eff} & \equiv & - \frac{n(n-1)}{2} \left[ \left(
\frac{\dot{b}}{b}\right)^2 + \frac{k_b}{b^2} \right] -
3n\frac{\dot{a}}{a} \frac{\dot{b}}{b} \, , \\
\bar{p}_{a,eff} & \equiv & n\frac{\ddot{b}}{b} + \frac{n(n-1)}{2}
\left[ \left( \frac{\dot{b}}{b}\right)^2 + \frac{k_b}{b^2} \right]
+ 2n \frac{\dot{a}}{a} \frac{\dot{b}}{b} \, , \\
\bar{p}_{b,eff} & \equiv & (n-1)\frac{\ddot{b}}{b} +
\frac{(n-1)(n-2)}{2} \left[ \left( \frac{\dot{b}}{b}\right)^2 +
\frac{k_b}{b^2} \right] + 3(n-1) \frac{\dot{a}}{a}
\frac{\dot{b}}{b} \, .
\end{eqnarray}
Note that an equation such as Eq.\ (\ref{Gjj eq with homog ED}) or
Eq.\ (\ref{Gjj eq with eff DE from ED}) is absent in the standard
cosmology, that is, a no-dark-geometry believer living in a
$(3+1)$-dimensional universe will never notice this equation,
while a higher-dimensional cosmologist will realize it as a
constraint equation for the behavior of the scale factor $b$ and
the pressure in the extra space $\bar{p}_b$.

Using equations of state $\bar{p}_a=w_a \bar{\rho}$ and
$\bar{p}_b=w_b \bar{\rho}$, we rearrange Eqs.\ (\ref{Gii eq with
homog ED}) and (\ref{Gjj eq with homog ED}) and obtain
\begin{eqnarray}
\frac{\ddot{a}}{a} &=&
-\left[\frac{(2n+1)-3(n-1)w_a+3nw_b}{n+2}\right]
\left(\frac{\dot{a}^2}{a^2}+\frac{k_a}{a^2}\right)
\nonumber \\
&& + \, \frac{n(n-1)}{2} \left[\frac{1+(n-1)w_a-nw_b}{n+2}\right]
\left(\frac{\dot{b}^2}{b^2}+\frac{k_b}{b^2}\right) \nonumber \\
&& - \, n \left[\frac{(n-1)-3(n-1)w_a+3nw_b}{n+2}\right]
\left(\frac{\dot{a}\dot{b}}{ab}\right) \, , \label{ddot a} \\
\frac{\ddot{b}}{b} &=& 3\left({\frac{1-3w_a+2w_b}{n+2}}\right)
\left(\frac{\dot{a}^2}{a^2}+\frac{k_a}{a^2}\right)  \nonumber \\
&& - \frac{n-1}{2} \left[\frac{(n+4)+3nw_a-2nw_b}{n+2}\right]
\left(\frac{\dot{b}^2}{b^2}+\frac{k_b}{b^2}\right) \nonumber \\
&& - 3\left(\frac{2+3nw_a-2nw_b}{n+2}\right)
\left(\frac{\dot{a}\dot{b}}{ab}\right) \, . \label{ddot b}
\end{eqnarray}

Before performing detailed numerical analyses, we can extract
essential features of the above equations and possible evolution
patterns governed by them via new variables
\begin{eqnarray}
u(t) & \equiv & \dot{a}/a  \, ,\\
v(t) & \equiv & \dot{b}/b  \; ,
\end{eqnarray}
and the parameter space expanded by
\begin{eqnarray}
X_a & \equiv & \frac{k_a}{a^2 u^2} \, , \\
X_b & \equiv & \frac{k_b}{b^2 u^2} \, , \\
 Y  & \equiv & v/u \, .
\end{eqnarray}
The use of the parameter space for analyzing the evolution of the
universe with extra dimensions was first suggested by Gu and Hwang
in \cite{Gu:2001ni,Gu:2002mz}. In terms of the above new
variables, Eqs.\ (\ref{G00 eq with homog ED}), (\ref{ddot a}), and
(\ref{ddot b}) can be rewritten as follows,
\begin{equation}
\left(1+X_a\right) + \frac{n(n-1)}{6}\left(Y^2+X_b\right) + nY =
\frac{8\pi\bar{G}\bar{\rho}}{3u^2} \, , %
\label{XY eq 00}
\end{equation}
\begin{eqnarray}
\lefteqn{\frac{1}{u^2}\left(\frac{\ddot{a}}{a}\right) =
\frac{\dot{u}}{u^2} + 1}
\nonumber \\
&=&  -\left[\frac{(2n+1)-3(n-1)w_a+3nw_b}{n+2}\right]
\left(1+X_a\right)
\nonumber \\
&& + \, \frac{n(n-1)}{2} \left[\frac{1+(n-1)w_a-nw_b}{n+2}\right]
\left(Y^2+X_b\right) \nonumber \\
&& - \, n \left[\frac{(n-1)-3(n-1)w_a+3nw_b}{n+2}\right] Y \, ,
\label{XY eq ddot a}
\end{eqnarray}
\begin{eqnarray}
\lefteqn{\frac{1}{u^2}\left(\frac{\ddot{b}}{b}\right) =
\frac{\dot{v}}{u^2} + Y^2}
\nonumber \\
&=& 3\left({\frac{1-3w_a+2w_b}{n+2}}\right)
\left(1+X_a\right)  \nonumber \\
&& - \frac{n-1}{2} \left[\frac{(n+4)+3nw_a-2nw_b}{n+2}\right]
\left(Y^2+X_b\right) \nonumber \\
&& - 3\left(\frac{2+3nw_a-2nw_b}{n+2}\right) Y \, . \label{XY eq
ddot b}
\end{eqnarray}
The above three equations indicate that the essential quantities
$\bar{G}\bar{\rho}$, $\ddot{a}/a$ and $\ddot{b}/b$, in unit of
$u^2$, are all functions of $X_a$, $X_b$ and $Y$ for given values
of $n$, $w_a$ and $w_b$. That is, $X_a$, $X_b$ and $Y$ form a set
of variables that can specify the state of a universe with extra
dimensions, provided that the unit of time is set to be the Hubble
time $u^{-1}$ at each moment. With the use of the parameter space
$(X_a \, , X_b \, , Y)$ we can plot regions for $\bar{\rho}>0$,
$\bar{\rho}=0$, or $\bar{\rho}<0$, and regions for $\ddot{a}>0$,
$\ddot{a}=0$, or $\ddot{a}<0$, in order to catch a basic picture
of the state of the universe at each position in the parameter
space. In addition, drawing flow vectors in the parameter space is
useful for sketching possible evolution patterns. Flow vectors in
the parameter space are vector fields:
\begin{equation}
\frac{d(X_a \, , X_b \, , Y)}{dt} = u \left[
-2X_a\left(1+\frac{\dot{u}}{u^2}\right) ,
-2X_b\left(Y+\frac{\dot{u}}{u^2}\right) ,
\left(\frac{\dot{v}}{u^2}-Y\frac{\dot{u}}{u^2}\right) \right] \, ,
\label{flow vector}
\end{equation}
which, in unit of $u$, are also functions of $(X_a \, , X_b \, ,
Y)$ for given values of $(n \, , w_a \, , w_b)$, as guaranteed by
Eqs.\ (\ref{XY eq 00})--(\ref{XY eq ddot b}). Accordingly the use
of the parameter space $(X_a \, , X_b \, , Y)$ makes it possible
to study the evolution of the universe before doing brute-force
numerical calculations.

In the rest of this paper we will only consider the universe with
flat extra dimensions, i.e., $k_b=0$, for simplicity. Accordingly,
in the next section we will make use of a two-dimensional
parameter space expanded by $X_a$ and $Y$ (i.e.\ the plane of
$X_b=0$) to study the evolution of the universe in various
possible eras, from the very early time to the present. [Hereafter
we will use $X$ instead of $X_a$ for convenience.] %
We will discuss three possible eras: 
the blazing era, the radiation-dominated era, and the
matter-dominated era, following the time order. The existence of
the blazing era is speculative. In this era the temperature of the
universe is so high that particles move relativistically in both
ordinary dimensions and extra dimensions and hence
$w_a=w_b=1/(3+n)$. In the expanding and cooling process of the
universe, after some time the temperature will be lower than the
inverse of the size of extra dimensions so that the motion in
extra dimensions, i.e.\ KK modes, cannot be excited and particles
are basically at the ground state and therefore pressureless in
extra dimensions thereafter.  After that time if the temperature
is high enough such that dominant particles are still relativistic
in ordinary dimensions, we have $w_a=1/3$ and $w_b=0$, as to be
designated as ``radiation-dominated era'' in this paper.
Conversely, if the temperature is so low in the later time that
the universe is dominated by particles that are nonrelativistic in
both ordinary dimensions and extra dimensions, we have
$w_a=w_b=0$, as to be designated as ``matter-dominated era''.

\section{The Evolution in Three Eras}

\subsection{The Blazing Era}%

\begin{figure}[h]
\centerline{\psfig{figure=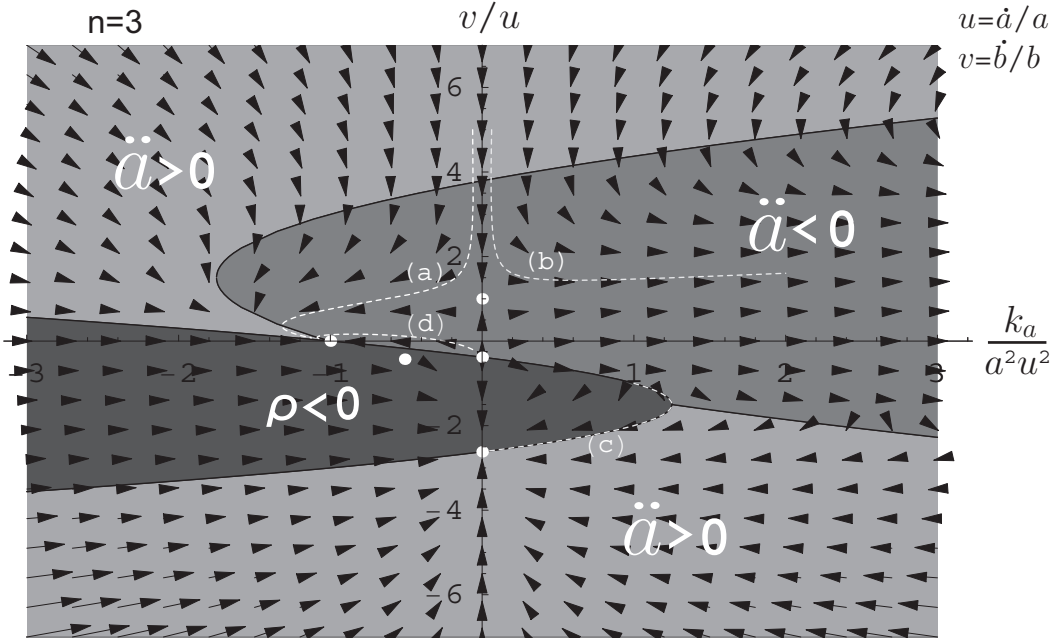,width=5.5in}} 
\caption{The $(v/u \, ,k_a/a^2u^2)$ parameter space describes
various states and possible evolution patterns of a blazing
universe, where the number of extra dimensions is specified to be
three for demonstration. Flow vectors and four evolution
trajectories (white dashed curves) are plotted.}
\label{fig:blazing}
\end{figure}

In the blazing era, $w_a=w_b=1/(3+n)$.
In Fig.\ \ref{fig:blazing} we plot regions for $\bar{\rho}>0$,
$\bar{\rho}=0$ or $\bar{\rho}<0$, which are obtained from Eq.\
(\ref{XY eq 00}), regions for $\ddot{a}>0$, $\ddot{a}=0$ or
$\ddot{a}<0$ from Eq.\ (\ref{XY eq ddot a}), flow vectors $d(X \,
, Y)/(udt)$ from Eqs.\ (\ref{XY eq ddot a})--(\ref{flow vector})
(assuming $u>0$), and four trajectories from numerical solutions,
where the number of extra dimensions $n$ has been specified to be
three for demonstration. The black area denoted by
``$\bar{\rho}<0$'' is a forbidden region if the energy density is
required to be positive. White dots stand for fixed points where
flow vectors are zero both in $X$ and $Y$ directions. The four
evolution trajectories correspond to numerical solutions of Eqs.\
(\ref{ddot a}) and (\ref{ddot b}) with the initial conditions
$(k_a/a^2u^2 , v/u)$ = (a) $(-0.0625,6.25)$, (b) $(0.0625,0.75)$,
(c) $(1,-0.9)$, and (d) $(-0.01,-0.1)$ respectively, representing
four possible evolution patterns:\footnote{Note that for all the
three eras we will only present some among all the possible
evolution patterns because many others are apparently not suitable
for describing our universe.}
\begin{enumerate}
\item[(a)] {\bf acceleration $\rightarrow$ deceleration
  $\rightarrow$
      acceleration}, eventually approaching asymptotically the attractor
      at $(-1,0)$ with stable extra dimensions and zero acceleration,
      possessing negative spatial curvature.
\item[(b)] {\bf acceleration $\rightarrow$ deceleration},
possessing
      increasing positive curvature contribution.
\item[(c)] {\bf deceleration $\rightarrow$ acceleration},
eventually
      approaching asymptotically the attractor at
      $\left\{ 0 \, ,-\left[ 3+\sqrt{3(n+2)/n}\right] / (n-1)) \right\}$
      with collapsing extra dimensions, possessing decreasing positive
      curvature contribution in the late time.
\item[(d)] {\bf deceleration $\rightarrow$ acceleration},
eventually
      approaching asymptotically the attractor at ($-$1, 0) with
      stable extra dimensions and zero acceleration, possessing
      negative spatial curvature.
\end{enumerate}

It is essential to ask whether in this model the inflation can
take place in this era without introducing the inflaton field.
From Fig.\ \ref{fig:blazing} we conclude that the universe cannot
have a sufficient inflating expansion, which makes the universe
nearly flat, in the blazing era unless the extra space is
collapsing violently or the expansion rate of the extra space is
much larger than that of the ordinary space (i.e.\ $v \gg u$)
initially. On the contrary, if the inflation occurs before the
blazing era such that the resultant curvature contribution is
extremely small, the universe may evolve nearly along the
$v/u$-axis during the blazing era. In this case, for a significant
part of initial states in the parameter space (more precisely, for
$v > \left[ -3+\sqrt{3(n+2)/n} \right] / (n-1)$ if $k_a/a^2u^2
\sim 0$ as required by the inflation) the universe tends to evolve
to the state at $(k_a/a^2u^2 \, , v/u) = (0,1)$, that is, the
ordinary space and the extra space tend to synchronize their
expansion rates. This feature provides a natural initial
condition, $k_a/a^2u^2 \sim 0$ and $v/u \simeq 1$, for the
succeeding radiation-dominated era.

\subsection{The Radiation-dominated Era}%
In the radiation-dominated era, $w_a=1/3$ and $w_b=0$. Fig.\
\ref{fig:radiation} shows regions corresponding to different signs
of $\bar{\rho}$ and $\ddot{a}$, flow vectors, four fixed points
(denoted by four white dots), and three evolution trajectories in
the parameter space (where $n$ is specified to be three for
demonstration). The three trajectories are with respect to initial
conditions $(k_a/a^2u^2 , v/u)$ = (a) $(-0.0625,5)$, (b)
$(0.0625,5)$, (c) $(1,-0.9)$, representing three possible
evolution patterns:

\begin{figure}[!h]
\centerline{\psfig{figure=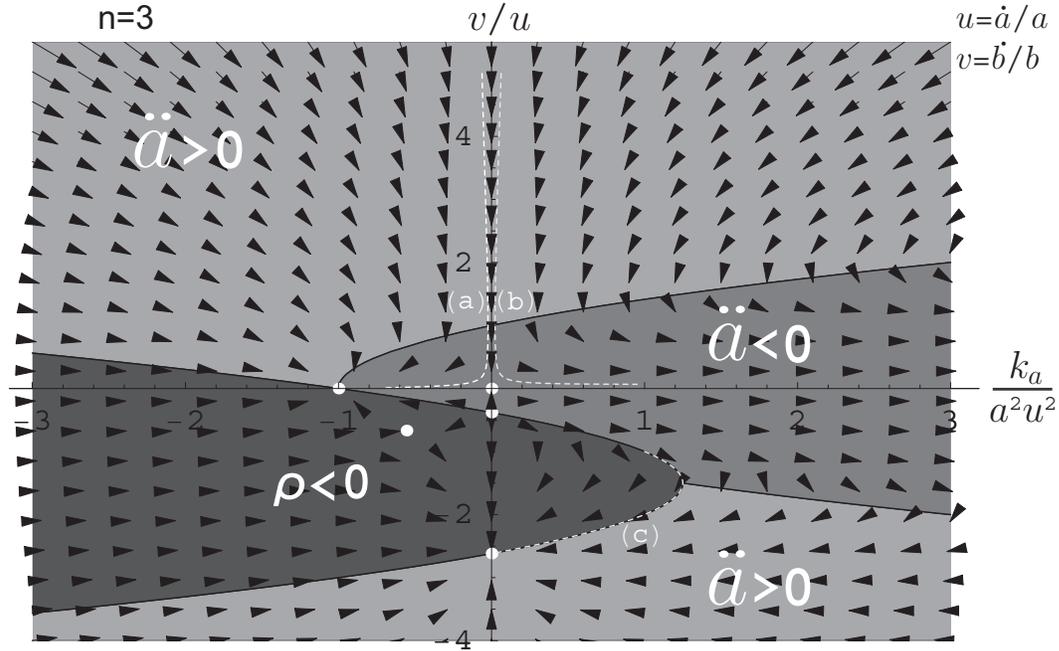,width=5.5in}} 
\caption{The $(v/u \, ,k_a/a^2u^2)$ parameter space describes
various states and possible evolution patterns of a
radiation-dominated universe, where the number of extra dimensions
is specified to be three for demonstration. Flow vectors and three
evolution trajectories (white dashed curves) are plotted.}
\label{fig:radiation}
\end{figure}

\begin{enumerate}
\item[(a)] {\bf acceleration $\rightarrow$ deceleration},
eventually
      approaching asymptotically the attractor at $(-1,0)$ with
      stable extra dimensions and zero acceleration, possessing
      negative spatial curvature.
\item[(b)] {\bf acceleration $\rightarrow$ deceleration},
eventually
      approaching asymptotically the $k_a/a^2u^2$-axis with stable
      extra dimensions, possessing increasing positive curvature
      contribution.
\item[(c)] {\bf deceleration $\rightarrow$ acceleration},
eventually
      approaching asymptotically the attractor at
      $\left\{ 0 \, ,-\left[ 3+\sqrt{3(n+2)/n}\right] / (n-1)) \right\}$
      with collapsing extra dimensions, possessing decreasing positive
      curvature contribution in the late time.
\end{enumerate}

The answer to the question regarding whether the inflation can be
generated in this radiation-dominated era without introducing the
inflaton field is similar to that in the blazing era. We conclude
from Fig.\ \ref{fig:radiation} that the universe cannot have a
sufficient inflation, which makes the universe flat enough, in
this era unless the extra space is collapsing violently or the
expansion rate of the extra space is much larger than that of the
ordinary space (i.e.\ $v \gg u$) initially.

An important feature in this era is that for about half of initial
conditions in the parameter space the corresponding evolution will
eventually approach to the states with stable extra dimensions,
i.e., $v/u \rightarrow 0$. This implies that the extra dimensions
have a significant possibility to be stabilized automatically
without resorting to additional artificial mechanisms.  For a
negligibly small expansion rate of extra dimensions $v$, we
can recover the standard cosmology 
as follows: With negligible contributions from $v$, zero curvature
in the extra space ($k_b=0$), and equations of state for this era
($\bar{p}_a=1/3 \bar{\rho}$ and $\bar{p}_b=0$), Eqs.\ (\ref{G00 eq
with eff DE from ED})--(\ref{Gjj eq with eff DE from ED}) become
\begin{eqnarray}
3 \left[ \left( \frac{\dot{a}}{a}\right)^2 + \frac{k_a}{a^2}
\right] &=& 8 \pi \bar{G} \bar{\rho} \, ,
\label{G00 eq with negligible eff DE from ED} \\
2 \frac{\ddot{a}}{a} + \left[ \left( \frac{\dot{a}}{a} \right)^2 +
\frac{k_a}{a^2} \right] &=& - 8 \pi \bar{G} \bar{\rho} / 3 \, ,
\label{Gii eq with negligible eff DE from ED} \\
3 \frac{\ddot{a}}{a} + 3 \left[ \left( \frac{\dot{a}}{a} \right)^2
+ \frac{k_a}{a^2} \right] &=& 0 \, , \label{Gjj eq with negligible
eff DE from ED}
\end{eqnarray}
where Eqs.\ (\ref{G00 eq with negligible eff DE from ED}) and
(\ref{Gii eq with negligible eff DE from ED}) are exactly the
Einstein equations in the standard cosmology and Eq.\ (\ref{Gjj eq
with negligible eff DE from ED}) is a redundant equation that can
be derived from the other two equations. [Indeed the fact that one
of the above equations is redundant indicates the existence of the
solution with $v=0$.] Consequently, with small enough $v$ in this
model we can preserve the essential predictions in the standard
cosmology, in particular the primordial abundance of light
elements produced during Big Bang Nucleosynthesis (BBN), in the
radiation-dominated era. In addition, this feature provides a
natural initial condition, $k_a/a^2u^2 \sim 0$ (guaranteed by
inflation presumably) and $v/u \simeq 0$, for the succeeding
matter-dominated era.

\subsection{The Matter-dominated Era}%

\begin{figure}[!h]
\centerline{\psfig{figure=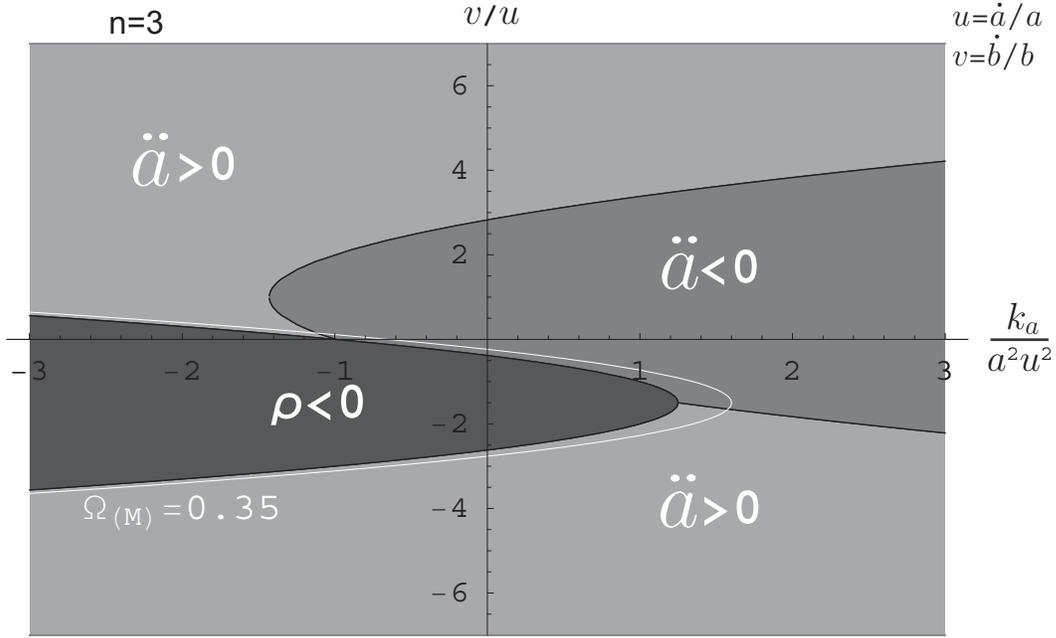,width=5.5in}} 
\caption{Various conditions in the $(v/u \, ,k_a/a^2u^2)$
parameter space for a matter-dominated universe are demonstrated.}
\label{fig:matter:conditions}
\end{figure}

\begin{figure}[!h]
\centerline{\psfig{figure=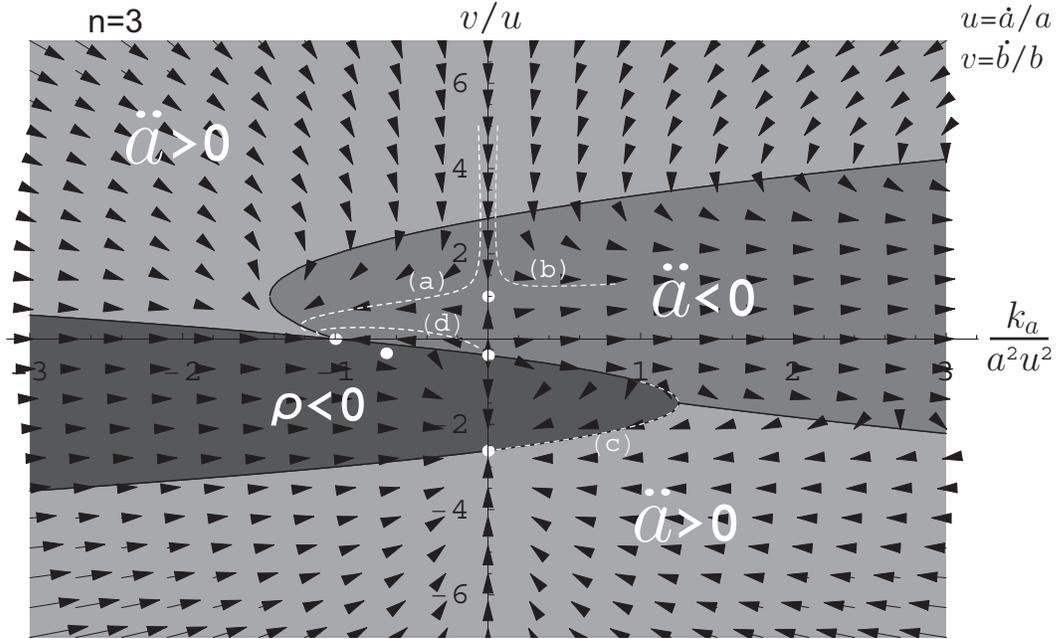,width=5.5in}} 
\caption{The $(v/u \, ,k_a/a^2u^2)$ parameter space describes
various states and possible evolution patterns of a
matter-dominated universe. Flow vectors and four evolution
trajectories (white dashed curves) are plotted.}
\label{fig:matter:flow}
\end{figure}

The evolution in the matter-dominated era, especially the possible
accelerating expansion of the present universe, is the main issue
that we will use more length to discuss in this paper. In this era
the universe is dominated by pressureless matter (without
introducing dark energy), therefore $\bar{\rho}=\bar{\rho}_M$ and
$w_a=w_b=0$. For demonstration the number of extra dimensions $n$
is chosen to be three for the plots in the parameter space. Fig.\
\ref{fig:matter:conditions} shows regions corresponding to various
signs of $\bar{\rho}_{(M)}$ and $\ddot{a}$ and the observational
bound to the energy density of the matter (mainly the cold dark
matter), which is represented by a solid curve denoted by
$\Omega_{(M)} = 0.35$.\footnote{The observational bound to the
energy density of the pressureless matter is $0.15 \lsim \Omega_M
\lsim 0.45$ \cite{Hagiwara:fs}. In our model, $\Omega_M \equiv
8\pi\bar{G}\bar{\rho}_{(M)}/3u_0^2$, where $u_0$ is the present
value of the Hubble parameter.} Fig.~\ref{fig:matter:flow} shows
flow vectors, four fixed points (denoted by four white dots), and
four evolution trajectories in the parameter space. The four
trajectories are with respect to initial conditions $(k_a/a^2u^2 ,
v/u)$ = (a) $(-0.0625,5)$, (b) $(0.0625,5)$, (c) $(1,-0.9)$, (d)
$(-0.0625,-0.1575)$, representing four possible evolution patterns
(that are somewhat similar to those in the blazing era):
\begin{enumerate}
\item[(a)] {\bf acceleration $\rightarrow$ deceleration
$\rightarrow$
      acceleration}, eventually approaching asymptotically the attractor
      at $(-1,0)$ with stable extra dimensions and zero acceleration,
      possessing negative spatial curvature.
\item[(b)] {\bf acceleration $\rightarrow$ deceleration},
possessing
      increasing positive curvature contribution.
\item[(c)] {\bf deceleration $\rightarrow$ acceleration},
eventually
      approaching asymptotically the attractor at
      $\left\{ 0 \, ,-\left[ 3+\sqrt{3(n+2)/n}\right] / (n-1)) \right\}$
      with collapsing extra dimensions, possessing decreasing positive
      curvature contribution in the late time.
\item[(d)] {\bf deceleration $\rightarrow$ acceleration},
eventually
      approaching asymptotically the attractor at ($-$1, 0) with
      stable extra dimensions and zero acceleration, possessing
      negative spatial curvature.
\end{enumerate}


The possible state and the allowed evolution pattern of the
universe are constrained by a variety of observations, such as
anisotropies of the cosmic microwave background (CMB), large-scale
structure (LLS), distance measurements of type Ia supernovae (SNe
Ia) etc. In addition to the limit to the variation of the
Newtonian gravitational constant $G_N$ that entails the a general
difficulty for this scenario \cite{Cline:2002mw}, we will consider
constraints on the ``acceleration parameter'' $\tilde{q}_0 \equiv
\ddot{a}_0/a_0 u_0^2$ (i.e.\ $-q_0$) and other essential
cosmological parameters such as the pressureless matter density
parameter $\Omega_{M}$ and the Hubble constant $H_0$ (i.e.\ $u_0$)
(where the subscript `0' denotes the present time).

As summarized in \cite{Uzan:2002vq}, the bound to $| \dot{G}_N/G_N
|$ ranges from $10^{-12}$ yr$^{-1}$ to $10^{-10}$ yr$^{-1}$ for
different observations. In the model with extra dimensions
considered here, the Newtonian gravitational constant varies along
with the variation of the size of extra dimensions in the way that
\begin{equation}
G_N = \bar{G}/V_n \propto b^{-n} \, ,
\end{equation}
where $V_n$ is the volume of the extra space. Accordingly,
\begin{equation}
\dot{G}_N / G_N = - n v \, .
\end{equation}
Consequently, the upper bound to $|v_0|$ is
(10$^{-12}$--10$^{-10}$)$\, \times \, n^{-1}$ yr$^{-1}$, and the
upper bound to $|(v/u)_0|$ is about (0.01--1)$ \, \times \, n^{-1}
h^{-1}$ (where we have used $u_0 = 100$ h km s$^{-1}$ Mpc$^{-1}$).
(For $h=0.71$ as suggested by WMAP \cite{WMAP2003} and the special
case with $n=3$, the upper bound ranges from 0.0047 to 0.47.)
Therefore the present state of the universe should be around the
$k_a/a^2u^2$-axis in the parameter space.

The above constraint on $|(v/u)_0|$, the requirement
$\ddot{a}_0>0$ from SN Ia data, and $0.15 \lsim \Omega_M \lsim
0.45$ \cite{Hagiwara:fs} together imply that the present state of
the universe should be in the up-left direction of the fixed point
at $(-1,0)$ in the parameter space (Fig.\
\ref{fig:matter:conditions}), and its evolution should follow
type-(a) or type-(d) trajectories in Fig.\ \ref{fig:matter:flow}.
Type-(d) trajectories are much more favored than type (a) for two
reasons: Firstly, type-(d) trajectories possess smaller $|v/u|$
through this era so that they are more consistent with the bound
to the variation of the Newtonian gravitational constant.
Secondly, it has been shown in the previous discussions that in
the radiation-dominated era the universe may approach to the
states with stable extra dimensions (i.e.\ $v/u \rightarrow 0$),
which should be the most probable initial conditions for the
succeeding matter-dominated era, accordingly leading to type-(d)
trajectories.

\begin{figure}[!h]
\centerline{\psfig{figure=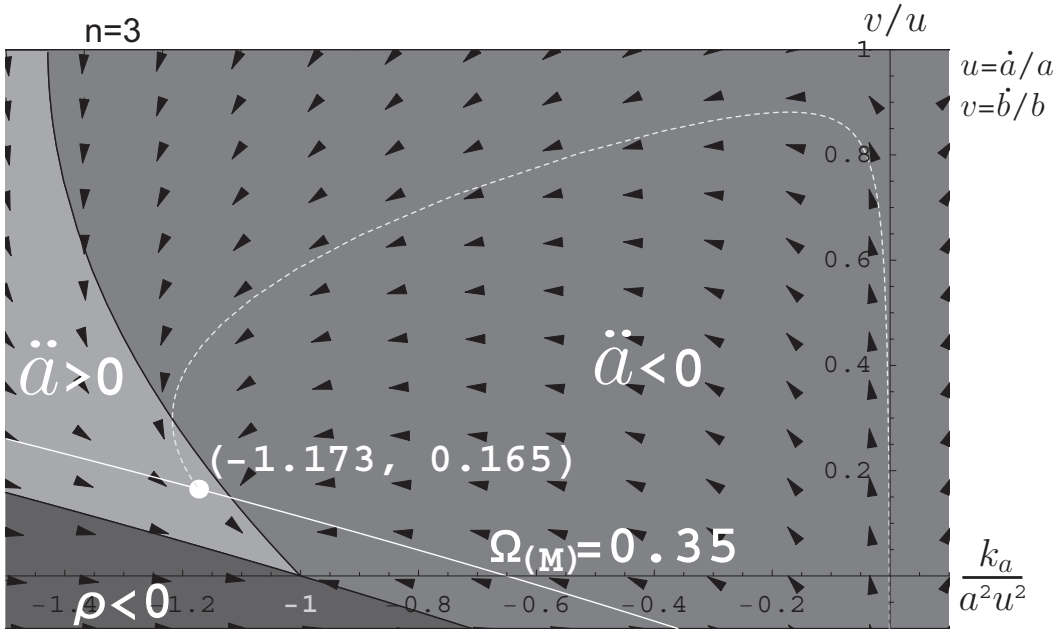,width=5.5in}} 
\caption{An enlarged plot for the region where (d)-type
trajectories pass. The trajectory (white dashed curve) which
entails the largest acceleration parameter $\tilde{q}_0$ is
plotted.} \label{fig:matter:zoomin}
\end{figure}

There is a boundary between type-(a) and type-(d) trajectories.
The intersection of this boundary and the present condition $0.15
\lsim \Omega_{(M)} \lsim 0.45$ in the parameter space provides an
upper bound, which is about $6.09 \times 10^{-2}$, to the
acceleration parameter $\tilde{q}_0$. These facts are summarized
in Fig.\ \ref{fig:matter:zoomin}, where the point $(-1.173,0.165)$
is the intersection of the curve $\Omega_{(M)}=0.35$ and the
boundary between type-(a) and type-(d) trajectories, and the white
dashed line denotes one of type-(d) trajectories which pass
through the neighborhood of this intersection point. In
particular, we note that the maximal-acceleration-parameter state
represented by this intersecting point entails small enough $v/u$
($v/u=0.165$) so that a moderate bound to the variation of the
Newtonian gravitational constant is satisfied. This fact indicates
that the essential difficulty from the variation of the Newtonian
gravitational constant can be somewhat avoided for type-(d)
trajectories. Nevertheless, this value of $\tilde{q}_0$ is smaller
than that required by SN Ia data together with the results of CMB
and LSS observations within the framework of the standard
cosmology \cite{Perlmutter:1999jt}. According to the result in
\cite{Perlmutter:1999jt}, the most probable values of
$(\Omega_M,w_X)$ corresponding to the requirement $\tilde{q}_0 =
6.09 \times 10^{-2}$ is about $(0.35,-0.51)$, which is roughly on
the margin of the $2.5 \sigma$ confidence interval. (In
\cite{Perlmutter:1999jt} $w_X$ denotes the equation of state of
dark energy.)

\begin{figure}[h!]
\centerline{\psfig{figure=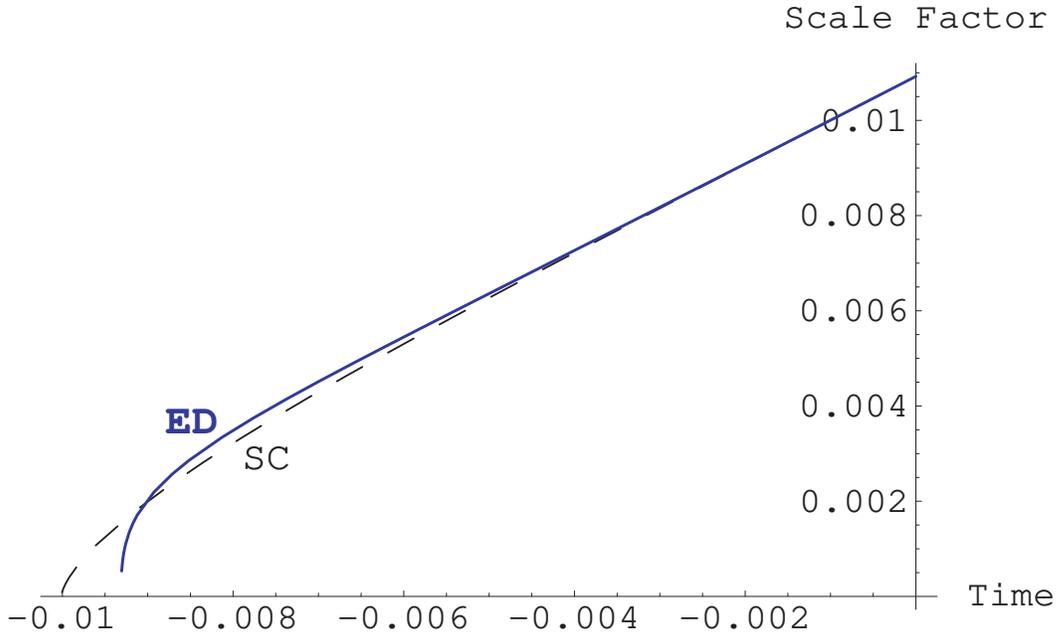,width=5.5in}} 
\caption{The evolution of the scale factor $a(t)$ for trajectories
in the ED and the SC model. } \label{fig:ED&SC evolution paths}
\end{figure}

Now we want to compare the trajectory in Fig.\
\ref{fig:matter:zoomin}, which entails the maximal value of
$\tilde{q}_0$ among type-(d) trajectories in our model with extra
dimensions, to be called ``ED model'', with that predicted in the
standard cosmology in which a flat universe is considered to
consist of pressureless matter with the density parameter
$\Omega_M$ and dark energy with the density parameter $\Omega_X$
($=1-\Omega_M$) and a constant state parameter $w_X$ ($w_X \equiv
p_X / \rho_X$), to be called ``SC model''.
In the ED model we need to specify the present condition about the
essential parameters $\Omega_{(M)}$, $X_0$, $Y_0$, and $u_0$ for
obtaining an evolution path, i.e., a numerical solution of the
Einstein equations (\ref{G00 eq with homog ED})--(\ref{Gjj eq with
homog ED}), while in the SC model the values of $\Omega_M$,
$w_{X}$, and $H_0$ ($H_0 = u_0$ by definition) are necessary
information. In order to make these two evolution paths to be as
similar as possible for the present time, we choose the present
conditions in the following consistent way: First of all the
acceleration parameter $\tilde{q}_0$ in both models is specified
to be $6.09 \times 10^{-2}$, i.e., the maximal value of
$\tilde{q}_0$ for type-(d) trajectories in the ED model. As
mentioned in the last paragraph about the ED model, the
intersection of the curve $\Omega_{(M)}=0.35$ and the (white
dashed) trajectory in Fig.\ \ref{fig:matter:zoomin} is at
$(-1.173,0.165)$, which provides the required consistent value of
$(X_0,Y_0)$. In summary, we choose the following present
conditions: $\Omega_{(M)}=0.35$, $X_0=-1.173$ and $Y_0=0.165$ for
the ED model and $\Omega_M=0.35$ and $\Omega_X=-0.51$ for the SC
model, as well as $H_0 = u_0 = 71$ km s$^{-1}$ Mpc$^{-1}$ from
WMAP \cite{WMAP2003}. Numerical solutions are shown in Fig.\
\ref{fig:ED&SC evolution paths}. These two evolution paths have
similar behavior at the present time but diverge in the earlier
time. In particular, the ED path entails a little shorter age.
Possible observational signatures from these discrepancies in the
age and the earlier-time behavior are under investigation.

\section{Conclusions and Discussions}
We have discussed the evolution of a universe under the influence
of flat extra dimensions in various eras via a well-chosen
parameter space that provides a systematic, useful, and convenient
way for analysis. Since we assume the flatness (zero curvature) of
the extra space, the only ingredient of extra dimensions that can
affect the evolution of the universe is the evolution of their
size. We have studied general features of the evolution, thereby
exploring whether the picture is viable, subject to constraints
from observations, for describing our world.

Through the above studies we have found a natural evolution
pattern from the blazing era to the matter-dominated era in this
model. In the blazing era, the ordinary space and the extra space
tend to synchronize their expansion rates, i.e., $v/u \rightarrow
1$, and meanwhile the universe is decelerating, presuming that an
extremely tiny curvature contribution $k_a/a^2 u^2$ has been
guaranteed initially by inflation.\footnote{This tendency
corresponds to the fixed point at $(k_a/a^2 u^2 \, , \,
v/u)=(0,1)$ in the parameter space for the blazing era, which is
stable in the Y direction but unstable in the X direction.} This
tendency provides a natural initial condition, $(k_a/a^2 u^2 \, ,
\, v/u)=(\pm \epsilon ,1)$, for the succeeding radiation-dominated
era (where `$\pm \epsilon$' indicates a tiny curvature
contribution).

In the radiation-dominated era, the expansion rate of the extra
space, $v$, tends to approach zero from $v/u \simeq 1$ (as
suggested above), meanwhile maintaining the decelerating
phase.\footnote{This tendency corresponds to the fixed point at
$(k_a/a^2 u^2 \, , \, v/u)=(0,0)$ in the parameter space for the
radiation-dominated era, which is stable in the Y direction but
unstable in the X direction.} It is a good feature for building
models with extra dimensions that the expansion rate of the extra
space decreases to zero automatically, that is, extra dimensions
are stabilized automatically in this era without resorting to any
artificial mechanism. In particular, for small enough $v$ we can
recover the standard cosmology (without extra dimensions) and the
essential predictions therein (in particular, Big Bang
Nucleosynthesis) in the radiation-dominated era. Therefore in our
model with extra dimensions the predictions in the standard
cosmology in the radiation-dominated era can be preserved in a
natural way.

In the matter-dominated era, a natural initial condition,
$(k_a/a^2 u^2 \, , \, v/u)=(\pm \epsilon , 0)$, is provided by the
preceding, radiation-dominated era. For the case of a negative
curvature, i.e.\ with the initial condition $(-\epsilon , 0)$, the
universe will eventually change its initial decelerating phase to
the accelerating one and approach the attractor at $(k_a/a^2 u^2
\, , \, v/u)=(-1,0)$ with stable extra dimension and significant
negative curvature contribution. As we have shown, this evolution
pattern is marginally consistent with the limit on the variation
of the Newtonian gravitational constant. The maximal acceleration
associated with this evolution pattern deviates from the one
required by SN Ia data together with CMB and LLS observations
(within the framework of the standard cosmology)
\cite{Perlmutter:1999jt} with $2.5 \sigma$ deviation. Thus this
evolution pattern is still not ruled out by the present
observations.

We note that in this natural evolution pattern the expansion rate
of extra dimensions will eventually approach zero, consequently
extra dimensions being stabilized automatically without resorting
to any artificial mechanism, in both the radiation-dominated and
the matter-dominated era. This is the key ingredient for
generating accelerating expansion of the present universe as well
as satisfying the limit on the variation of the Newtonian
gravitational constant (both caused by the evolution of extra
dimensions), meanwhile preserving essential predictions in the
standard cosmology for the radiation-dominated era. In particular,
we emphasize that the naturalness of this evolution pattern, whose
behavior is not sensitive to the initial conditions, indicates a
solution to the cosmic coincidence problem (``why now'' problem)
of dark energy %
--- why dark energy starts dominating the universe now, or, more
precisely, why the accelerating phase starts at the present epoch. %
The existence of this natural evolution pattern implies that the
late-time accelerating expansion of an open universe
\footnote{Actually this open universe is nearly flat in the
earlier time.} in the matter-dominated era is guaranteed,
accordingly the cosmic coincidence problem being solved, in our
flat ED model with the help of inflation in the very early time.

Nevertheless, type-(d) trajectories in general entail a too short
age of the universe 
and negative curvature in the ordinary space rather than almost
zero curvature suggested by CMB data within the framework of
standard cosmology. These features might eventually rule out this
evolution pattern for describing our universe. Further detailed
studies about the consistency with observations are in progress.
Moreover, in order to complete the picture of the scenario for a
universe with extra dimensions and find the best-fit trajectory
therein, a more general model with nonzero curvature in the extra
space is also under investigation.

\section*{Acknowledgements}
This work is supported in part through the Taiwan CosPA Project,
as funded by the Ministry of Education (MoE 89-N-FA01-1-3), and
also supported by the National Science Council, Taiwan, R.O.C.
(NSC 92-2112-M-002-051).


\end{document}